\documentclass[useAMS,usenatbib]{mn2e}

\newcommand{\Msun}{\hbox{$\hbox{M}_\odot\;$}}
\newcommand{\Rsun}{\hbox{$\hbox{R}_\odot\;$}}
\newcommand{\kms}{\hbox{${\rm km}\:{\rm s}^{-1}\;$}}
\newcommand{\Msuno}{\hbox{$\hbox{M}_\odot$}}

\usepackage{graphicx}
\usepackage{natbib}

\title[Fast orbital decays of black hole X-ray binaries: XTE J1118+480 and A0620--00]
{Fast orbital decays of black hole X-ray binaries: XTE
J1118+480 and A0620--00\footnotemark[\star]\thanks{Based on observations made 
with the Gran Telescopio Canarias (GTC), instaled in the Spanish
Observatorio del Roque de los Muchachos of the Instituto de
Astrof{\'\i}sica de Canarias, in the island of La Palma.}
}
\author[J.~I.~Gonz\'alez~Hern\'andez et al.]
{J.~I.~Gonz\'alez~Hern\'andez$^{1,2}$\thanks{E-mail:
jonay@iac.es; rrl@iac.es, jorge.casares@iac.es}, R. Rebolo$^{1,2,3}$ 
and J. Casares$^{1,2}$\\
$^{1}$Instituto de Astrof{\'\i }sica de Canarias (IAC), 
E-38205 La Laguna, Tenerife, Spain\\
$^{2}$Depto. Astrof{\'\i }sica, Universidad de La 
Laguna (ULL), E-38206 La Laguna, Tenerife, Spain
$^{3}$Consejo Superior de Investigaciones 
Cient{\'\i}ficas, Spain
}
\begin{document}

\date{Accepted 2013 October XX. Received 2013 May 15}

\pagerange{\pageref{firstpage}--\pageref{lastpage}} \pubyear{2013}

\maketitle

\label{firstpage}

\begin{abstract}
We present new 10.4m-GTC/OSIRIS spectroscopic observations of the 
black hole X-ray binary \mbox{XTE J1118+480} that confirm the 
orbital period decay at $\dot P=-1.90\pm0.57$~ms~yr$^{-1}$. 
This corresponds to a period change of $-0.88\pm0.27$~$\mu $s 
per orbital cycle. We have also collected observations of the black hole X-ray
binary \mbox{A0620--00} to derive an orbital period derivative of 
$\dot P=-0.60\pm0.08$~ms~yr$^{-1}$ ($-0.53\pm0.07$~$\mu $s/cycle). 
Angular momentum losses due to gravitational radiation are unable to explain 
these large orbital decays in these two short-period black hole binaries. 
The orbital period decay
measured in A0620--00 is very marginally consistent with the predictions of
conventional models including magnetic braking, although significant
mass loss ($\dot M_{\rm BH} / \dot M_2 \le 20\%$) from the system is required. 
The fast spiral-in of the star in \mbox{XTE J1118+480}, however, does not fit 
any standard model and may be 
driven by magnetic braking under extremely high magnetic fields and/or may 
require an unknown process or non-standard theories of gravity. 
This result may suggest an evolutionary sequence in which the orbital period 
decay begins to speed up as the orbital period decreases. 
This scenario may have an impact on the evolution and lifetime of black hole 
X-ray binaries.
\end{abstract}


\section{Introduction}

Black hole X-ray binaries (BHXBs) are ideal astrophysical sites 
to investigate angular momentum losses (AMLs), especially those driven
by magnetic braking~\citep[MB;][]{ver81}, gravitational 
radiation~\citep[GR;][]{lan62,tay82} 
and mass loss~\citep[ML;][]{rap82}. Magnetic braking is supposed to be the main
mechanism responsible for AMLs in very compact binaries with short orbital
periods of several hours~\citep{rap82}, but there is still a lot of debate on
the suitable prescription for MB~\citep{pod02a,iva06b,yun08b}. 
The measurement of period variations 
can help to understand these processes.
In particular, the BHXBs \mbox{XTE J1118+480} and \mbox{A0620--00} are very 
interesting systems because, since their discovery, they have been intensively 
studied and therefore their dynamical parameters are well constrained (see
Table~\ref{ttpar}).  
\mbox{XTE J1118+480} has an orbital period of 4.08~hr~\citep{tor04, gon12a} and 
a black hole mass in the range $M_{\rm BH}\sim$~6.9--8.2~\Msun~\citep{kha13}. 
\mbox{A0620--00} has a longer orbital period of 7.75~hr~\citep{mcc86} and 
similar black hole mass in the range $M_{\rm BH}\sim$~6.3--6.9
\Msun~\citep{can10}. 

\begin{table*}
\centering
\begin{minipage}{140mm}
\caption{Kinematical and dynamical binary parameters of 
\mbox{XTE J1118+480} and \mbox{A0620--00}}
\begin{tabular}{@{}lcccc@{}}
\hline
{Parameter} & {XTEJ1118$+$480} & {Ref.}$^\star$ & {A0620$-$00} & {Ref.}$^\star$ \\
\hline
$v \sin i$         & $96^{+3}_{-11}$~\kms         & [1] & $82\pm2$~\kms                & [7] \\
$i$                & $73.5\pm5.5$                 & [2] & $51.0\pm0.9$		       & [8] \\
$k_2$              & $708.8\pm1.4$~\kms           & [3] & $435.4\pm0.5$~\kms	       & [7] \\
$k_2$              & $710.0\pm2.6$~\kms           & [4] & --	       & -- \\
$q=M_2/M_{\rm BH}$ & $0.024\pm0.009$              & [1] & $0.060\pm0.004$	       & [7] \\
$f(M)$             & $6.27\pm0.04$~\Msun          & [3] & $2.762\pm0.009$~\Msun	       & [4] \\
$M_{\rm BH}$       & $7.46^{+0.34}_{-0.69}$~\Msun & [4] & $6.61^{+0.23}_{-0.17}$~\Msun & [4] \\
$M_2$              & $0.18\pm0.06$~\Msun          & [4] & $0.40\pm0.01$~\Msun	       & [4] \\
$a_c$              & $2.54\pm0.06$~\Rsun          & [4] & $3.79\pm0.04$~\Rsun          & [4,9] \\
$R_2$              & $0.34\pm0.05$~\Rsun          & [4] & $0.67\pm0.02$~\Rsun          & [4,9] \\
$P_{\rm orb,1}$    & $0.1699339(2)$~d             & [5] & $0.323014(4)$~d	       & [10] \\
$P_{\rm orb,2}$    & $0.1699338(5)$~d             & [6] & --                           & --  \\
$P_{\rm orb,3}$    & $0.1699337(2)$~d             & [4] & --                           & --  \\
$P_{\rm orb,0}$    & $0.16993404(5)$~d            & [4] & $0.32301415(7)$~d            & [4] \\
$T_0$              & $2451868.8921(2)$~d          & [4] & $2446082.6671(5)$~d	       & [4] \\
$\dot P_{\rm orb}$ & $-6.01\pm1.81 \times 10^{-11}$~s~s$^{-1}$ & [4] & $-1.90\pm0.26 \times 10^{-11}$~s~s$^{-1}$  & [4] \\
$\dot P_{\rm orb}$ & $-1.90\pm0.57$~ms~yr$^{-1}$  & [4] & $-0.60\pm0.08$~ms~yr$^{-1}$  & [4] \\
$\dot P_{\rm orb}$ & $-0.88\pm0.27$~$\mu$s~cycle$^{-1}$ & [4] & $-0.53\pm0.07$~$\mu$s~cycle$^{-1}$ & [4] \\
$\dot P_{\rm orb, MC,o}$ & $-1.98\pm0.56$~ms~yr$^{-1}$  & [4] & $-0.63\pm0.08$~ms~yr$^{-1}$  & [4] \\
$\dot P_{\rm orb, MC,c}$ & $-1.98\pm0.59$~ms~yr$^{-1}$  & [4] & $-0.62\pm0.12$~ms~yr$^{-1}$  & [4] \\
\hline
\end{tabular}
{\\
$^\star$~References: [1]~\citet{cal09}; [2]~\citet{kha13};
[3]~\citet{gon08b}; [4] This work; [5] ~\citet{tor04}; 
[6]~\citet{gon12a}; [7]~\citet{nei08}; 
[8]~\citet{can10}; [9]~\citet{gon11}; [10]~\citet{mcc86}
}
\end{minipage}
\label{ttpar}
\end{table*}

The standard theory~\citep[e.g.][]{ver93,pod02a,tay82}, predicts that the 
orbital period first derivative from AMLs due to MB and ML in these two BHXBs 
should be $\dot P_{\rm MB,ML} \sim -0.02$~ms~yr$^{-1}$ whereas GR accounts 
only for $\dot P_{\rm GR} \le -0.01$~ms~yr$^{-1}$, according to the dynamical 
parameters of these BHXBs. \citet{joh09a, joh09b} 
tried to measure the orbital period decay of these BHXBs but failed due to the 
limited baseline of the available observations. Recently,
\citet{gon12a} have been able to detect, for 
the first time, orbital period variations in a short period BHXB. They measured  
$\dot P \sim -1.8$~ms~yr$^{-1}$ in \mbox{XTE J1118+480}, i.e. several
order of magnitude larger than expected from conventional models of AMLs. These
authors suggested that extremely high magnetic fields in the secondary star at
about 10--30~kG could explain the fast spiral-in of the star to the black hole
in this system. In the current work we confirm the value obtained in
\citet{gon12a} with new observations and present the detection of also a 
negative orbital period derivative in the BHXB \mbox{A0620--00}. 

\section{Observations}

\begin{figure}
\centering
\includegraphics[height=8.5cm,angle=90]{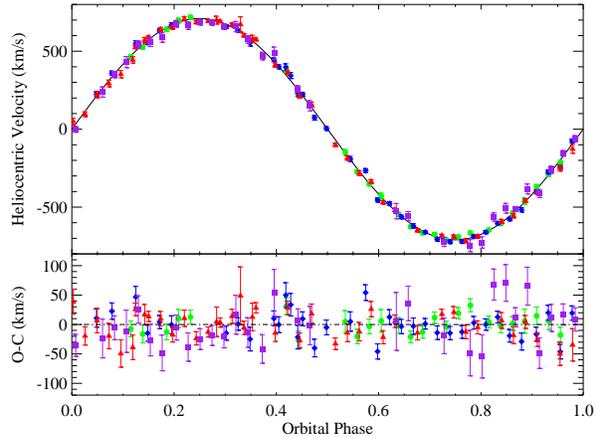}
\caption{\scriptsize{{\it Top panel}: radial velocities of the 
secondary star in \mbox{XTE J1118+480} 
obtained from the previous GTC/OSIRIS spectroscopic data taken 
on the three nights of January 7 (green filled circles), February 8 
(blue filled diamonds) and April 25 (red filled triangles) in 2011, 
and the new GTC/OSIRIS data acquired on UT 2012 January 12, 
folded on the best-fitting orbital solution. 
{\it Bottom panel}: residuals of the fit, with a rms of $\sim 20 
{\rm km}\ {\rm s}^{-1}$.}}
\label{frv}   
\end{figure}

We have conducted new spectroscopic observations of 
\mbox{XTE J1118+480} using the 10.4m Gran Telescopio Canarias (GTC) 
equipped with the OSIRIS spectrograph~\citep{cep00} at the 
Observatorio del Roque de los Muchachos in La Palma 
(Canary Islands, Spain). Thirty four new medium-resolution spectra
($\lambda/\delta\lambda\sim2,500$) were obtained on 2012 
January 12 UT. Radial velocity (RV)
measurements (see Fig.~\ref{frv}) were extracted from every spectrum 
as in~\citet{gon12a}. The orbital period in the new RV curve 
was fixed to new value, $P_{\rm orb}=0.1699337(2)$~d, which was derived
with all GTC/OSIRIS RV points, and over a baseline of one year,
including three nights in 2011 and one in 2012.
This orbital period measurement is 17~ms shorter although still consistent 
with that derived by \citet{tor04}.
In Table~\ref{ttpar} we list the updated kinematical and dynamical
parameters of these BHXBs.

\begin{table}
\centering
\begin{minipage}{80mm}
\caption{Time at inferior conjunction of the secondary star 
in \mbox{XTE J1118+480}.}
\begin{tabular}{@{}lccc@{}}
\hline
{$N$} & {$T_n-2450000$}\footnote{Times in HJD of the $n^{th}$ 
inferior conjunction, $T_n$, of \mbox{XTE J1118+480}, and 
uncertainties, $\delta T_n$.} & 
$\delta T_n$ & Refs.\footnote{[1]~\citet{wag01}; [2]~\citet{tor04}; [3]
~\citet{zur02}; [4]~\citet{gon08b}; [5]~\citet{gon12a} [6]~This work.} \\
\hline
0     & 1868.8916  & 0.0004 & [1] \\
66    & 1880.1086  & 0.0004 & [2] \\
904   & 2022.5122\footnote{From photometric measurements.} & 0.0004 & [3] \\
6950  & 3049.93347 & 0.00007 & [4] \\
21772 & 5568.6936  & 0.0003 & [5] \\
21960 & 5600.6413  & 0.0002 & [5] \\
22407 & 5676.6017  & 0.0002 & [5] \\
23949 & 5938.6395  & 0.0005 & [6] \\
\hline
\end{tabular}
\end{minipage}
\label{ttnxte}
\end{table}

\begin{table}
\centering
\begin{minipage}{80mm}
\caption{Time at inferior conjunction of the secondary star 
in \mbox{A0620--000}.}
\begin{tabular}{@{}lccc@{}}
\hline
$N$ & $T_n-2440000$\footnote{Times in HJD of the $n^{th}$ 
inferior conjunction, $T_n$, of \mbox{A0620--00}, and 
uncertainties, $\delta T_n$.} 
& $\delta T_n$ & Refs.\footnote{[1]~\citet{mcc86}; 
[2]~This work: these $T_n$ values have been corrected from times 
at maximum velocity to times at orbital phase 0 using the orbital
period in~\citet{mcc86}; [3]~\citet{joh09a};  
[4]~\citet{gon10}; [5]~\citet{sha04}; [6]~\citet{nei08}} \\
\hline
0     &  6082.6673  & 0.0008  & [1,2] \\
6764  &  8267.5347  & 0.0002  & [2,3] \\
17957 & 11883.0313  & 0.0002  & [4] \\
20321 & 12646.6365  & 0.0005  & [5] \\
24773 & 14084.69485 & 0.00005 & [6] \\
\hline
\end{tabular}
\end{minipage}
\label{ttna06}
\end{table}

\section{Orbital period decay}

\begin{figure}
\centering
\includegraphics[height=8.5cm,angle=90]{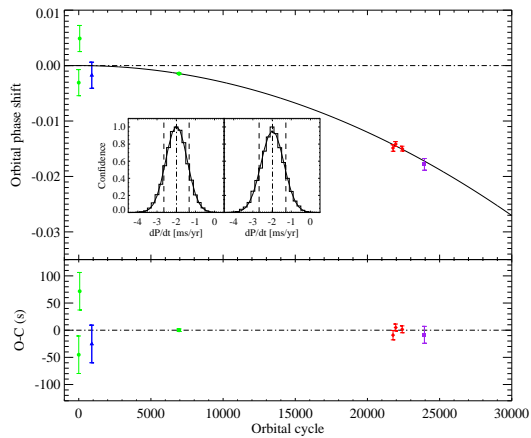}
\caption{\scriptsize{{\it Top panel}: orbital phase shift at the time of the 
inferior conjunction (orbital phase 0), $T_n$, of the secondary star in the 
BHXB \mbox{XTE J1118+480} versus the orbital 
cycle number, $n$, folded on the best-fit parabolic fit. 
The error bars give the uncertainties $\delta T_n$ (see
text and Table~\ref{ttnxte}).
Green filled circles are spectroscopic determinations, 
the blue filled triangle is a photometric measurement, 
red filled diamonds, and the new violet square are GTC/OSIRIS 
spectroscopic determinations (see Table~\ref{ttnxte}).
The small panels show two MonteCarlo (MC) simulations of
10,000 realizations taking into account the uncertainties of each $T_n$ point: 
(i) with the observed data set, i.e. using $T_n$ values as a center of the 
MC distributions (left small panel), and (ii) using the points on the 
parabolic fit (rigth small panel).
{\it Bottom panel}: residuals of the fit of the $T_n$ values versus 
the cycle number $n$.}} 
\label{fphxte}   
\end{figure}

The new spectroscopic data of \mbox{XTE 1118+480} have been 
used to derive one new value, $T_n$, of the inferior conjunction of 
the secondary star in this system (see Table~\ref{ttnxte}). 
Assuming a constant rate of change of the orbital period, the time, 
$T_n$, of the $n$th orbital cycle can be expressed 
as $T_n=T_0+P_0 n+\frac{1}{2} P_0 \dot P n^2$, where $P_0$ is the 
orbital period at time $T_0$ of the reference cycle ($n=0$), 
$\dot P$ is the orbital period time derivative, and $n$, 
the orbital cycle number. We use the IDL routine {\scshape curvefit} 
to perform a parabolic fit with a $\chi_\nu^2\sim1.4$ and obtain a 
period derivative of $\dot P = -(5.87\pm2.05) \times 10^{-11}$~s/s. 
A linear fit ($\dot P =0$; $\chi_\nu^2 \sim 2.5$) 
and a third-order polynomial fit (including $\ddot P$; $\chi_\nu^2 \sim 1.7$) 
provide worse results. 
We perform an F-test to evaluate how well the parabolic
fit reproduces this set of data with respect to the linear and cubic fit, 
and obtain significance values of 0.05 for the first-order versus second-order 
polynomial, and 0.87 for the second-order versus the third-order polynomial 
fit, indicating that the second-order polynomial provides a better 
representation of the current set of data. However, we note  
the low scatter of the last five $T_n$ around the parabolic fit which may
indicate that the error bars are overestimated. We therefore scale down the 
error bars of these five $T_n$ values by a factor $\sim0.4$ so to get 
$\chi_\nu^2\sim1$ for these five points, from which we obtain a 
period derivative determination of $\dot P = -(6.01\pm1.81) \times
10^{-11}$~s/s. 
In Fig.~\ref{fphxte} we have depicted the orbital phase
shift, defined as $\phi_n=\frac{T_n-T_0}{P_0}-n$, 
of each of the $T_n$ values as a function of the orbital cycle 
number $n$, together with the best-fit second-order solution.     
This figure shows a clear deviation from the null variation and 
that  $\dot P$ is negative. Our result, which can be expressed as 
$\dot P=-1.90\pm0.57$~ms~yr$^{-1}$, shows a determination at 
the 3.3$\sigma$ level (i.e. at $\sim 99$\% confidence).
We perform two MonteCarlo (MC) simulations: (i) with the 
observed $T_n$ points by randomly varying their values with the
the uncertainties $\delta T_n$ in a normal distribution; 
(ii) with the fitted $T_n$ points on the parabolic fit.
We fit 10,000 realizations of the simulated $T_n$ points: for case (i) 
with weights in the fit given as $1/(\delta T_n)^2$; 
and for case (ii) without weights.
The resulting histograms of $\dot P$ are shown in two small panels 
within Fig.~\ref{fphxte}.
These MC simulations provides value of
$\dot P_{\rm MC}=-1.98\pm 0.58$~ms~yr$^{-1}$ and confirms 
the orbital shrinkage in \mbox{XTE J1118+480} (see 
Table~\ref{ttpar}).

\begin{figure}
\centering
\includegraphics[height=8.5cm,angle=90]{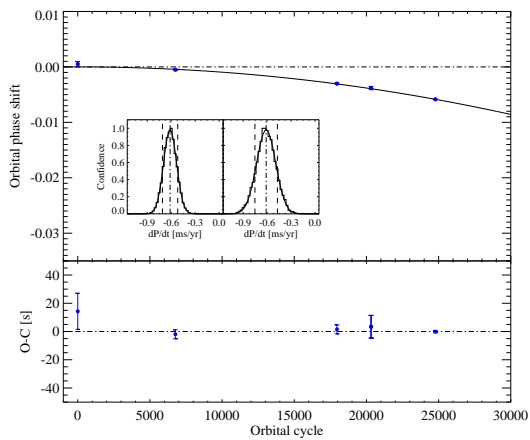}
\caption{\scriptsize{Same as Fig.~\ref{fphxte} but for the 
orbital decay in the BHXB \mbox{A0620--00}. 
All data points (filled circles) are spectroscopic determinations.}} 
\label{fpha06}   
\end{figure}

Previous attempts to estimate the orbital period derivative in 
\mbox{A0620--00} probably failed due to the small number of 
observations~\citep{joh09a}. We have added one additional and very
precise measurement of $T_0$ from \citet{gon10}.  
Thus, we also perform a fit to the $T_n$ times of \mbox{A0620--00} (see
Table~\ref{ttna06}),by scaling the error bars (by a factor $\sim0.2$) 
until we get $\chi_\nu^2\sim1$, allowing us to obtain 
$\dot P = -(1.90\pm 0.26) \times 10^{-11}$~s/s.
In Fig.~\ref{fpha06} we display the data points, including the point in
\citet{gon10}, together with the best fit. This BHXB  also shows a  
negative orbital period derivative, $\dot P=-0.60\pm0.08$~ms~yr$^{-1}$. 
The MC simulation confirms the orbital decay in \mbox{A0620--00} 
(see Table~\ref{ttpar}).

\section{Discussion and conclusions}

These BHXBs exhibit negative orbital period derivative but at different 
rate; \mbox{XTE J1118+480} with a shorter orbital period ($P_{\rm
orb}\sim 4.08$~hr) has a faster orbital decay than \mbox{A0620--00} 
($P_{\rm orb}\sim7.75$~hr), both faster than predicted by conventional
models of AMLs due to GR, MB and ML. In fact, standard theory foresees a 
$\sim$1.4 faster orbital decay for \mbox{A0620--00}.
This suggests an evolutionary sequence in which the orbital decays 
observed in short-period BHXBs tend to be faster as the 
companion star approaches the black hole.
 
\citet{gon12a} provided a theoretical estimate of the orbital period 
derivative, $\dot P_{\rm MB,ML}$, when considering only AMLs due to MB and ML 
using the expressions given in~\citet{joh09a}. They demonstrated 
that $\dot P_{\rm MB,ML}$ cannot account for the fast orbital decay observed 
in \mbox{XTE J1118+480}. Standard prescriptions of MB, 
assuming $\gamma =2.5$~\citep{ver93} and adopting arbitrarily 
$\beta=-\dot M_{\rm BH}/\dot M_2=0.5$ \citep{pod02a}, and the specific angular
momentum, $j_w=1$, carried away by the mass lost from the system, provide
values of $\dot P_{\rm MB,ML} \sim -0.024$ and $-0.017$~ms~yr$^{-1}$ for 
\mbox{A0620--00} and \mbox{XTE J1118+480}, respectively. 
Only when considering, in the extreme and probably unrealistic case, the 
values $\gamma =0$, i.e the strongest possible MB effect, and 
$\beta=0$, i.e all the mass transferred by the secondary star is lost from 
the system, we get 
$\dot P_{\rm MB,ML} \sim -0.27$ and $-0.86$~ms~yr$^{-1}$ for 
\mbox{A0620--00} and \mbox{XTE J1118+480}, which are consistent at 1$\sigma$
and $2\sigma$ with the observed measurements, respectively.

At orbital periods shorter than 3~hr, standard theory predicts that 
gravitational radiation begins to dominate, but for these systems, AMLs 
due to MB is 30-80 times larger than those caused by GR, according to the
dynamical parameters of these BHXB (see Table~\ref{ttpar}). 
Only at $P_{\rm orb}<$~40~min, does
$\dot P_{\rm GR}$ rise to values between $-0.4$ and $-1.4$~ms~yr$^{-1}$,
comparable to the observed determinations of $\dot P$ in these BHXBs. 
At the current orbital
period, $\dot P_{\rm GR} \sim -0.006$ and $-0.009$~ms~yr$^{-1}$ for 
\mbox{A0620--00} and \mbox{XTE J1118+480}, respectively, i.e. GR provides a 
$\sim$1.5 faster orbital decay for \mbox{XTE J1118+480}, although not 
sufficient to explain the observations. 

\citet{gon12a} suggested that extremely high magnetic fields in the secondary
star may help to explain the fast orbital decay in \mbox{XTE J1118+480}. The
current estimates of magnetic fields at the surface of secondary stars, 
estimated from Equation (5) in \citet{jus06}, and assuming the mass lost 
by wind is equal to the mass transfer 
rate~\citep[derived from Equation (9) in][]{kin96b}, are $B_S \sim 2$-$30$~kG 
(at $\gamma =2.5$ and $\gamma =0$) for \mbox{XTE J1118+480} but 
$B_S \sim 0.4$-$1$~kG for~\mbox{A0620--00}. 
This may be connected with chromospheric activity,
induced by rotation, on the companion star of \mbox{A0620--00} proposed 
by~\citet{gon10} to explain the observed companion's H$\alpha$ emission.
These high magnetic fields are not common in
highly-rotating low-mass stars \citep[e.g.][]{pha09}. However, \citet{jus06}
propose that short period X-ray binaries could be originated from binaries 
with secondary stars as intermediate-mass magnetically 
peculiar Ap/Bp stars which typically have magnetic fields of $\sim 20$~kG.
\citet{has02} detected enhanced CNO-processed material 
in the spectra of the accretion disk of \mbox{XTE J1118+480} which indicates
that the donor star may descend from an intermediate-mass star with a 
mass of $M_2\sim1.5$~\Msuno. 
More recently, \citet{fra09} studied the evolution of this system and 
conclude that the donor mass right after the black hole formation should 
be in the range 1--1.6~\Msuno. The models proposed by \citet{jus06} begin with 
higher secondary masses of about 
3--5~\Msun and effective temperatures at the end of their evolutionary tracks 
that are too hot compared to the those measured in these BHXB secondary
stars~\citep{gon04a,gon06}. However, observations of Ap stars with high magnetic
fields of $\sim 15-25$~kG, such as HD~154708, BD+0$^\circ$~4535 and HD~178892,
have $T_{\rm eff} \sim 6800-7700$~K and masses 
1.5--1.8~\Msuno~\citep[e.g.][]{nes08,elk10,rya06} which may alleviate the
initial mass problem of these models. Therefore, the scenario proposed 
by~\citet{jus06} may provide an explanation for the fast orbital decay in 
these systems, if the secondary star can retain the high magnetic fields of 
several kG. 

Orbital period modulations of the order of $\Delta P/P\sim10^{-5}-10^{-6}$ with
different signs have been observed in many type of objects, e.g. 
in V471 Tau~\citep{ski88}, in cataclysmic variables~\citep[CVs][]{pri75,war88}, 
W UMa stars~\citep{glo86}, and RS CVn systems~\citep{hal80}. 
A variable gravitational quadrupole moment, due to internal deformations, 
produced by magnetic activity in the outer convection zone has been 
suggested as the mechanism responsible for those 
period changes~\citep{app87,app92}.
This mechanism also predict stellar luminosity variations which should be
modulated with the period of the stellar magnetic activity cycle. 
Long-term quasi-periodic modulations of the X-ray flux have also been
seen in the light curves of steadily accreting neutron star LMXBs, and 
interpreted as a consequence of stellar magnetic activity
cycles~\citep{kot10}. 
The orbital period derivative in BHXBs could be the result of a magnetic cycle 
modulation, yielding a subsurface magnetic field of $\sim 0.3$ and 
$0.1$~kG for \mbox{XTE J1118+480} and \mbox{A0620--00}, respectively, due 
to the transition mechanism~\citep{app92}. 
The distortion mechanism would require subsurface 
magnetic fields of $\sim 12.3$ and~$3.2$~kG for \mbox{XTE J1118+480} and 
\mbox{A0620--00}. However, these two BHXBs show large orbital period decays
which makes it quite unlike that what we are observing are 
period modulations. Possible future observations with a baseline of five to
ten years may answer this question.

\citet{joh09b} suggested that the orbital period decays in short-period
BHXBs can be used to set constraints on the rate at which black holes can
evaporate in the Anti-de Sitter (AdS) braneworld Randall-Sundrum gravity 
model~\citep{ran99} via the emission of a large number of conformal field 
theory (CFT) modes~\citep{emp03}. 
The dynamical parameters and orbital period derivative of 
\mbox{XTE J1118+480} provide a new very tight constraint on 
the AdS curvature radius of $L \le 18$~$\mu$m at 2$\sigma$, 
which is more restricted than the best current table-top experiment 
upper-limit of $L \le 44$~$\mu$m~\citep{kap07}. 
Finally, other alternative theories of gravity such as the scalar Gauss-Bonnet (sGB)
theory have been recently proposed as explanation for the fast orbital decays in
these BHXBs~\citep{yag12}. The possible future measurement of the second orbital
period derivative, $\ddot P$, will probably help to distinguish among these 
theories of gravity, in particular using future gravitional wave observations 
with space-borne interferometers.

\section*{Acknowledgments}
J.I.G.H. acknowledges financial support from the Spanish Ministry 
project MINECO AYA2011-29060, and also from the Spanish Ministry of Economy 
and Competitiveness (MINECO) under the 2011 Severo Ochoa Program 
MINECO SEV-2011-0187. We would like to thank the referee, Phil Charles, 
for his helpful comments which have certainly improved the paper.
We are grateful to T. Marsh for the use of the MOLLY analysis package.
This work has made use of the IRAF facilities.

\bibliographystyle{mn2e}
\bibliography{lmxbs}


\bsp

\label{lastpage}

\end{document}